\documentclass[14pt,a4paper]{article}
\usepackage[cp1251]{inputenc}
\usepackage{amsmath}
\usepackage{amsfonts}
\usepackage{amssymb}
\usepackage{rotating}
\usepackage{fancyhdr} 
\makeatletter

\usepackage[]{graphicx}

\usepackage{listings}

\usepackage{tocloft}
\usepackage{geometry}
\usepackage[font=Large, labelsep=endash]{caption}
\newlength\myindention
\DeclareCaptionFormat{myformat}%
{\hspace*{\myindention}#1#2#3}
\DeclareCaptionLabelSeparator{endash}{ --- }
\newcommand{\biggI}{\renewcommand{\baselinestretch}{1.3}}
\renewcommand{\@oddhead}{\hfil \normalsize Numerical Modeling of the Internal Temperature in the Mammary Gland \hfil}
\renewcommand{\@evenhead}{\hfil \normalsize Polyakov M.V. et al. \hfil}
\makeatother

\makeatletter
\renewcommand{\@oddfoot}{\hfil \normalsize \arabic{page}\hfil}
\renewcommand{\@evenfoot}{\hfil \normalsize \arabic{page}\hfil}
\makeatother

\fontencoding{T1}
\fontfamily{pcr}
\selectfont

\makeatletter
 \renewcommand*\l@section[2]{%
   \ifnum \c@tocdepth >\z@
     \addpenalty\@secpenalty
     \addvspace{0em \@plus\p@}%
     \setlength\@tempdima{1.5em}%
     \begingroup
       \parindent \z@ \rightskip \@pnumwidth
       \parfillskip -\@pnumwidth
       \leavevmode
       \advance\leftskip\@tempdima
       \hskip -\leftskip
       #1\nobreak\hfil \nobreak\hb@xt@\@pnumwidth{\hss #2}\par
     \endgroup
   \fi}
 \makeatother

\lstdefinelanguage{cs}
  {morekeywords={abstract,event,new,struct,as,explicit,null,switch
		base,extern,object,this,bool,false,operator,throw,
		break,finally,out,true,byte,fixed,override,try,
		case,float,params,typeof,catch,for,private,uint,
		char,foreach,protected,ulong,checked,goto,public,unchecked,
		class,if,readonly,unsafe,const,implicit,ref,ushort,
		continue,in,return,using,decimal,int,sbyte,virtual,
		default,interface,sealed,volatile,delegate,internal,short,void,
		do,is,sizeof,while,double,lock,stackalloc,
		else,long,static,enum,namespace,string, },
	  sensitive=false,
	  morecomment=[l]{//},
	  morecomment=[s]{/*}{*/},
	  morestring=[b]",
}

\biggI

\begin{document}
\setcounter{page}{128}
\newcounter{save}\setcounter{save}{\value{section}}
{\def\addtocontents#1#2{}%
\def\addcontentsline#1#2#3{}%
\def\markboth#1#2{}%

\noindent
Polyakov M., Khoperskov A., Zamechnic T. Numerical Modeling of the Internal Temperature in the Mammary Gland //
Lecture Notes in Computer Science, 2017, v.10594, p.128-135

\noindent
https://doi.org/10.1007/978-3-319-69182-4\_14

\thispagestyle{empty}
\begin{center}
 \Large \textbf{Numerical Modeling of the Internal \\
 Temperature in the Mammary Gland}
\end{center}
\vspace{0.2cm}
\begin{center}
M.V. Polyakov$^1$, A.V. Khoperskov$^1$ and T.V. Zamechnic$^2$
\end{center}
\begin{center}
$^1$ Volgograd State University, Volgograd, Russia \\
{\ttfamily{\texttt{\{m.v.polyakov, khoperskov\}@volsu.ru}}} \\
$^2$Volgograd State Medical University, Volgograd, Russia \\
{\ttfamily{\texttt{tvzamechnic.61@mail}}}
\end{center}
}

{\large \textbf{Abstract.}  The microwave thermometry method for the diagnosis of breast cancer is based on an analysis of the internal temperature distribution.This paper is devoted to the construction of a mathematical model for increasing the accuracy of measuring the internal temperature of mammary glands, which are regarded as a complex combination of several components, such as fat tissue, muscle tissue, milk lobules, skin, blood flows, tumor tissue.
Each of these biocomponents is determined by its own set of physical parameters.
Our numerical model is designed to calculate the spatial distributions of the electric microwave field and the temperature inside the biological tissue.
We compare the numerical simulations results to the real medical measurements of the internal temperature.}

\section{Introduction}

{\large The method of microwave thermometry (RTM) has become significant recently for oncological diagnostics \cite{Akki-Arunachalam-2014,Barett,vesnin,vrba}. This method allows to conduct surveys frequently and without any harm to health \cite{shah}. The RTM is based on measuring the temperature inside the biological tissue at various points \cite{Avila-Castro,Bardati,Carr,novochadov}. Then the received data is subject to intellectual analysis to assist doctor with the diagnosis. Using the method of combined thermography, we consistently measure temperatures at 9 different points of the female breast \cite{losev,novochadov}.
Instrumental internal and surface temperatures of tissues are determined by the intensity of their thermal radiation in the microwave and infrared ranges, respectively.
Body temperature distribution is an important criterion in the diagnosis of diseases \cite{novochadov,shah} because it characteri~-~zes a person's functional state.
Tumors of the breast cause relatively high heat emission at an early stage, which leads to a local increase in the temperature of the tissues.

To increase the efficiency of diagnosis of oncological disease we study the results of a series of computer experiments on calculation the internal temperature in the microwave range with the radiothermometry method.
 An important problem of modeling biological tissues is the variability of individual physical characteristics of biological tissue of different people.
We use the model that is based on the numerical integration of Maxwell's equations and heat equation for multicomponent biological tissue.

The novelty of our approach is based on multifaceted computational models that take into account the multicomponent properties of biological tissues and their complex spatial structure.
As a rule, the heat sources associated with the blood flow are considered in the capillary approximation, when the heat release is evenly distributed over the volume of the tissue \cite{Gonzalez,vesnin}.
Our model contains a branched system of blood vessels of different diameters, and the heat release is not uniform.}
{\large
\begin{figure}[th]
\centering\includegraphics[width=0.8\hsize]{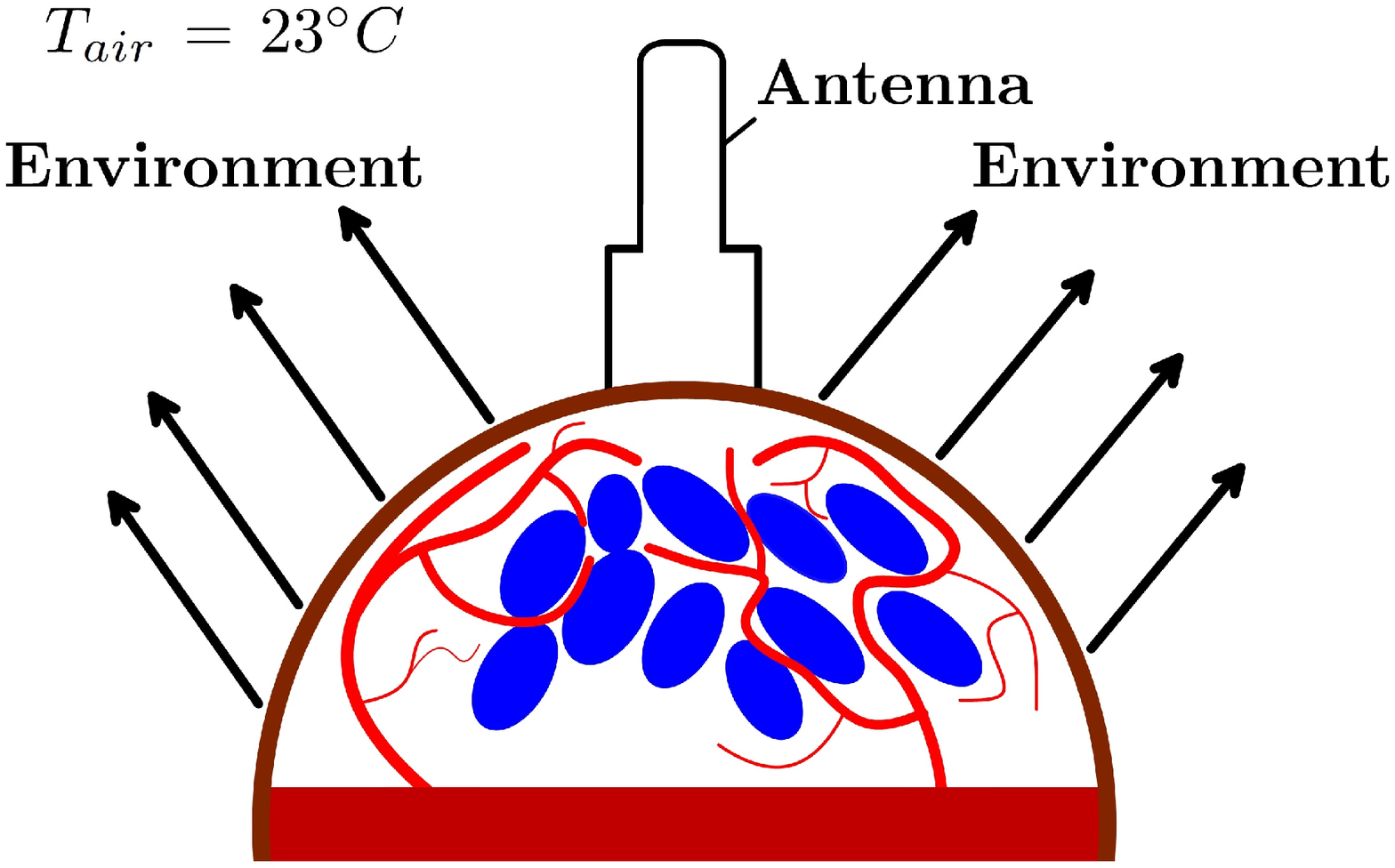}\vskip 0.05\hsize
\begin{center}
{\large Fig. 1. The general scheme of our model}
\end{center}
\label{Fig-model}
\end{figure}}

{\large The complex spatial structure of a mammary gland and small-scale inhomogeneity (Fig. 1) require the use of unstructured numerical grids for calculating electric and temperature fields.
Typical sizes of numerical cells are approximately $0.02-0.1$\,mm.}

\section{Mathematical model}

{\large Computer modeling is based on the numerical solving of the heat equation for biological tissue \cite{Avila-Castro,Gonzalez,losev,vesnin}:
\begin{equation}\label{eq-polykov-1}\displaystyle
\begin{array}{rcl}
   \displaystyle\rho({\bf r})c_{p}({\bf r})\frac{\partial T_{mod}}{\partial t}({\bf r},t)&=&\nabla(\delta({\bf r})\nabla T_{mod}({\bf r},t)) +
   \\ \ \\
   \ &+& Q_{bl}({\bf r},t)+Q_{met}({\bf r},t)-Q_{rad}({\bf r},t) \,,
\end{array}
\end{equation}
where $ \rho $ is the volume density of tissue, $c_{p}({\bf r}) $ is the specific heat of material,
$ T_{mod}$ is the simulated temperature, $ Q_{bl}$ is the heat source from blood vessels,
$ Q_{met}$ is the heat source from metabolic processes in biological tissues,
$ Q_{rad}$ is radiation cooling,
$\delta$ is the coefficient of thermal conductivity,
$\displaystyle \nabla=\left\{\frac{\partial}{\partial x},\frac{\partial}{\partial y}, \frac{\partial}{\partial z}\right\}$ is the nabla operator.}

{\large The intensity of heating is determined by the temperature difference between tissue $ T $ and blood $ T_{bl} $, and the specific heat of the blood $c_{p,b}$
\begin{equation}
   Q_{bl}=- \rho\rho_{bl}c_{p,b}\omega_{bl}(T-T_{bl}),
\end{equation}
where $ \omega_ {bl} $ is the intensity of blood flow in the heating region, which can vary over a wide range.

Heat exchange at the biological tissue boundary with the environment is defined as
\begin{equation}
({\bf n}\nabla T)=\frac{h_{air}}{\delta ({\bf r})}\, \big[ T({\bf r})-T_{air} \big]\,,
\end{equation}
where $ {\bf n} $ is a unit vector normal to the surface of the female breast.
The value of $ T_{air} $ characterizes the temperature of the environment and $ h_{air} $ is the heat transfer coefficient.

Stationary distribution of the electric field is constructed using the calculation for the establishment  and solving the nonstationary Maxwell's equations:
\begin{equation}\label{eq-Makswell}
    \frac{\partial {\bf B}}{\partial t} + {\rm rot}({\bf E}) = 0 \,,
\end{equation}
\begin{equation}\label{eq-Makswell}
     \frac{\partial {\bf D}}{\partial t} - {\rm rot}({\bf H}) = 0 \,,
\end{equation}
\begin{equation}\label{eq-Makswell}
 {\bf B}=\mu({\bf r})\, {\bf H}\,,\quad {\bf D}=\varepsilon({\bf r})\, {\bf E} \,.
\end{equation}}
{\large Antenna allows to measure thermal radiation in the frequency range $ f_{min} \leq f \leq f_{max} $. Biological tissue has an inhomogeneous temperature. This method gives a weighted average temperature $ T_{int} $ in the region $ V_0 $ \cite{Foster}. The error of the method is due to the noise temperature of the receiver $ T_{noise} $, the mismatch effects in the antenna $ s(f) $, the environmental effect $ T_{env} $. The internal temperature is determined by the integral representation \cite{polyakov,vesnin}:
\begin{equation}
T_{int}=\int\limits_{\Delta f}\left[(1-|s(f)|^{2}) \left(\int\limits_{V}W({\bf r},f)\, T_{mod}({\bf r})\,dV+T_{env} \right)+|s(f)|^{2}T_{noise}\right]\,df
\end{equation}
with weighting function
\begin{equation}\displaystyle
W({\bf r},f)=\frac{F({\bf r},f)}{\int\limits_{V}F({\bf r},f)\,dV}\,, \quad \int\limits_{V}W({\bf r},f)\,dV=1 \,,
\end{equation}

\begin{equation}
F({\bf r},f)=\frac{1}{2}\,\sigma({\bf r},f)\,|E({\bf r},f)|^{2} \,.
\end{equation}

The value $W$ according to the principle of reciprocity coincides with the density of the power $ F$ (W/m$^3$) and depends on the electrical conductivity of $ \sigma $ and electric field ${\bf E}$ inside a certain volume $ V $. Efficiency of temperature $ T_{int} $ measurementby a radiometer strongly depends on electromagnetic interference $T_{noise}$, which requires proper shielding.

The receiving antenna is the key element of the radiothermometer.}

{\large An important task is to achieve a low level of mismatch $s$ between the antenna and biotissue. The antenna should ensure a deep penetration into the biotissue. It depends on the ratio of energy from brown fat tissue to total power.}

\section{Results of computer modeling}
{\large The input data for modeling are the physical parameters of the biocomponent (Table~1).
\begin{center}
\textbf{Table 1.} Physical parameters of components \cite{losev}
\end{center}
\begin{table}[!h]
\centering
\label{parameters}
\begin{tabular}{|l|c|c|c|c|}
\hline
                              & Skin & \begin{tabular}[c]{@{}c@{}} Mammary \\ gland \end{tabular} & \begin{tabular}[c]{@{}c@{}}Connective \\         tissue\end{tabular} & Bloodstream \\ \hline
\begin{tabular}[c]{@{}c@{}} Dielectric \\ permeability, $\varepsilon$ \end{tabular} & 53.5---57.2 & 5.1---5.9    & 44.2---48.1                                                                      & 1.54---1.96     \\ \hline
\begin{tabular}[c]{@{}c@{}}Electric \\ conduction, \\ $\sigma$ (1$/$Ohm) \end{tabular}           & 0.92---1.31 & 0.03---0.09            & 2.19---2.68                                                                    & 45.8---48.2     \\ \hline
\begin{tabular}[c]{@{}c@{}} Resistivity,\\ $\cal R$ (Ohm$\cdot$m)\end{tabular}       & 53.1---56.9   & 13.9---15.7              & 1.31---1.74                                                                     & 1.22---1.7      \\ \hline
\end{tabular}
\end{table}

\begin{figure}
\centering{
\includegraphics[width=0.6\hsize]{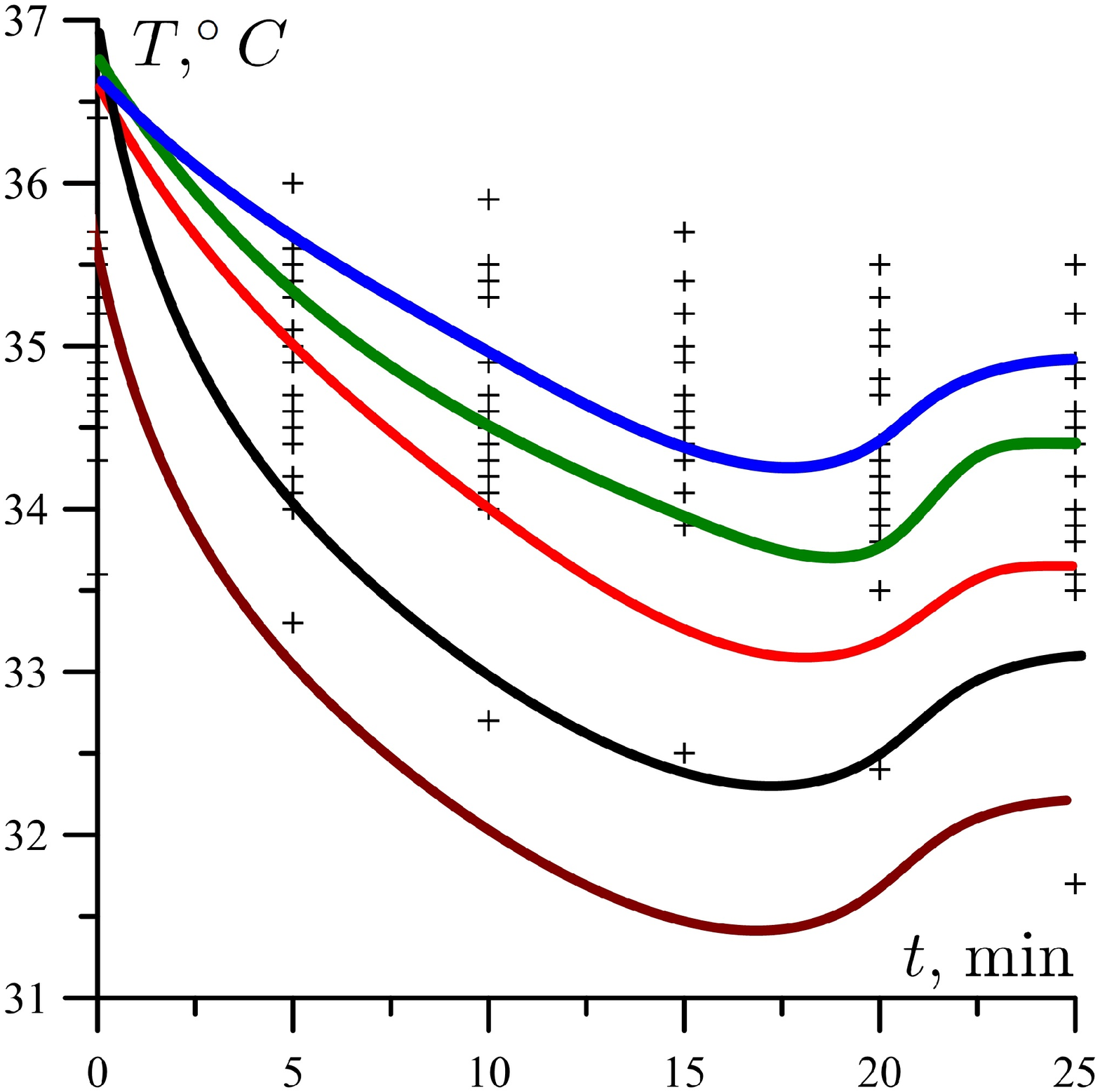}\vskip 0.02\hsize
}
\begin{center}
{\large Fig. 2. Comparison of the results of numerical experiments (color lines) with the data of real medical measurements (+)}
\end{center}
\label{Fig-graph2}
\end{figure}

The $ T_{mod}$ calculation is performed using the Comsol Multiphysics version~4.3\,a \cite{Datta-Comsol}. The $ \vec {E} $ calculation is performed using the CST Microwave Studio simulation package \cite{Kurushin-CST}. Calculation of the deep temperature is carried out with the mathematical package Scilab.

The mechanism of adaptation of a living organism to environmental conditions deserves special attention.
Physiological adaptation is associated with the regulation of the physiological functions of the organism.
Thermoregulation is the ability of living organisms to maintain the temperature of their body at a constant level or to change it within certain limits.}

{\large
\begin{figure}[!ht]
\centering{
\includegraphics[width=6cm]{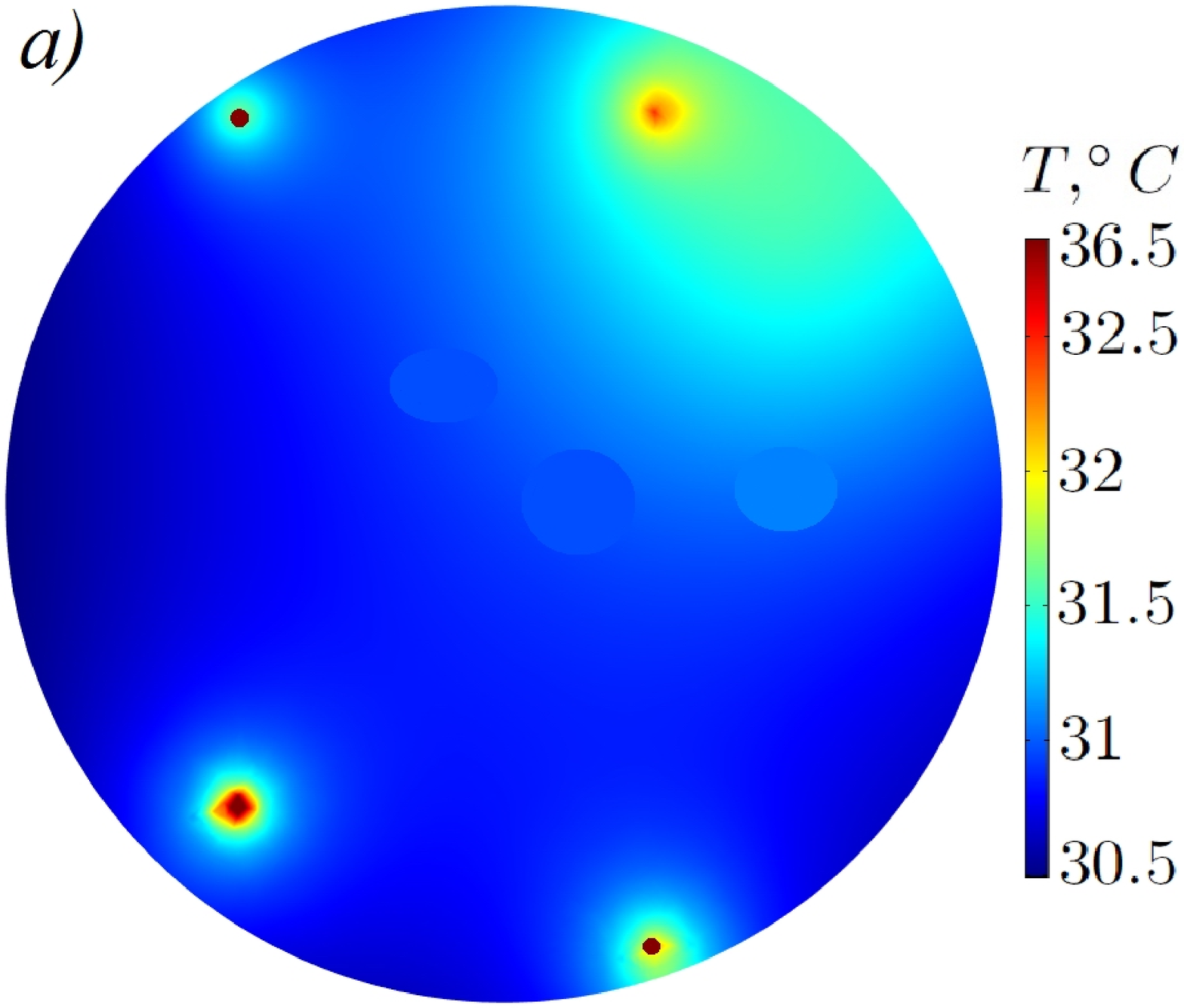}
\includegraphics[width=6cm]{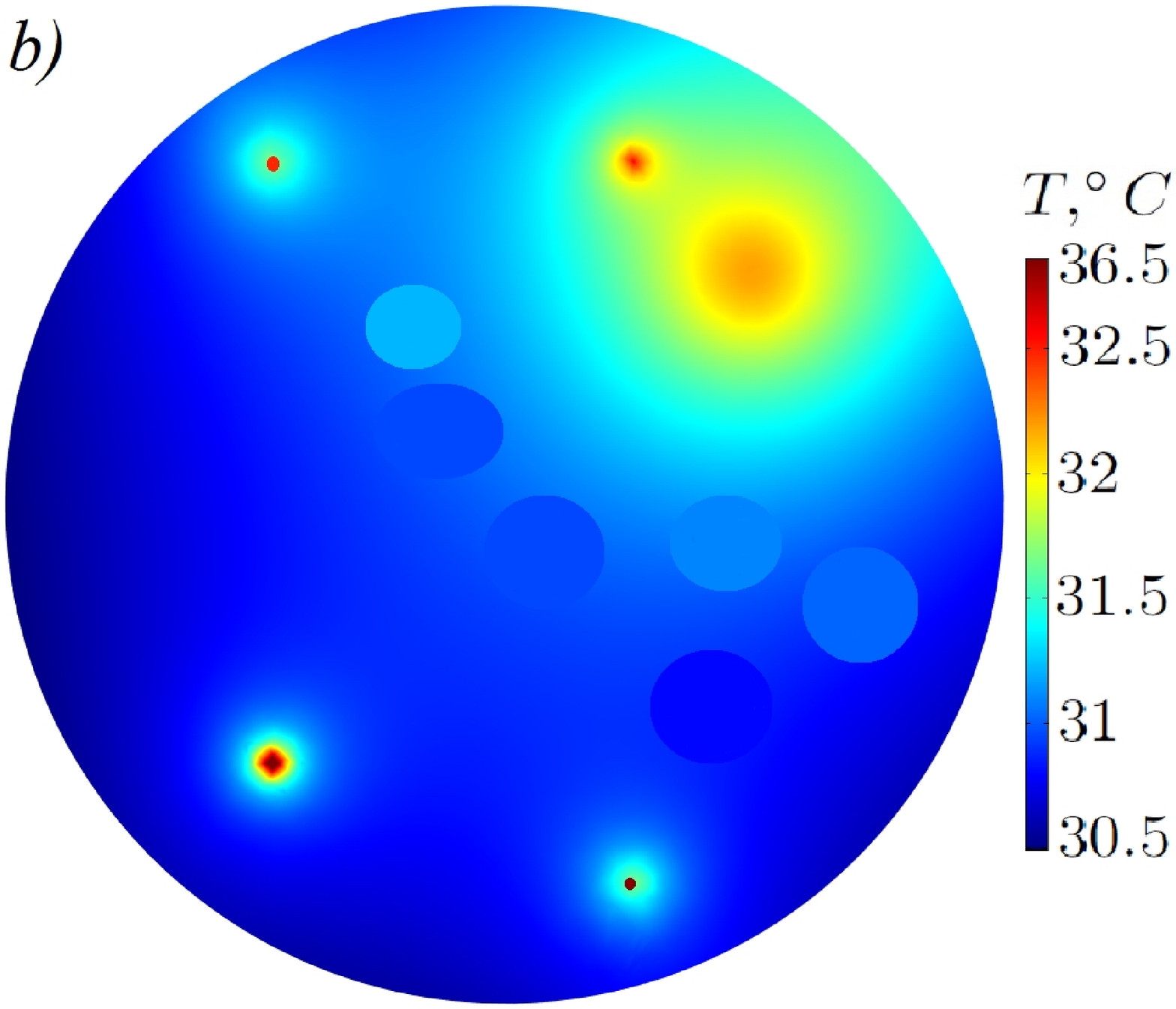}
\includegraphics[width=6cm]{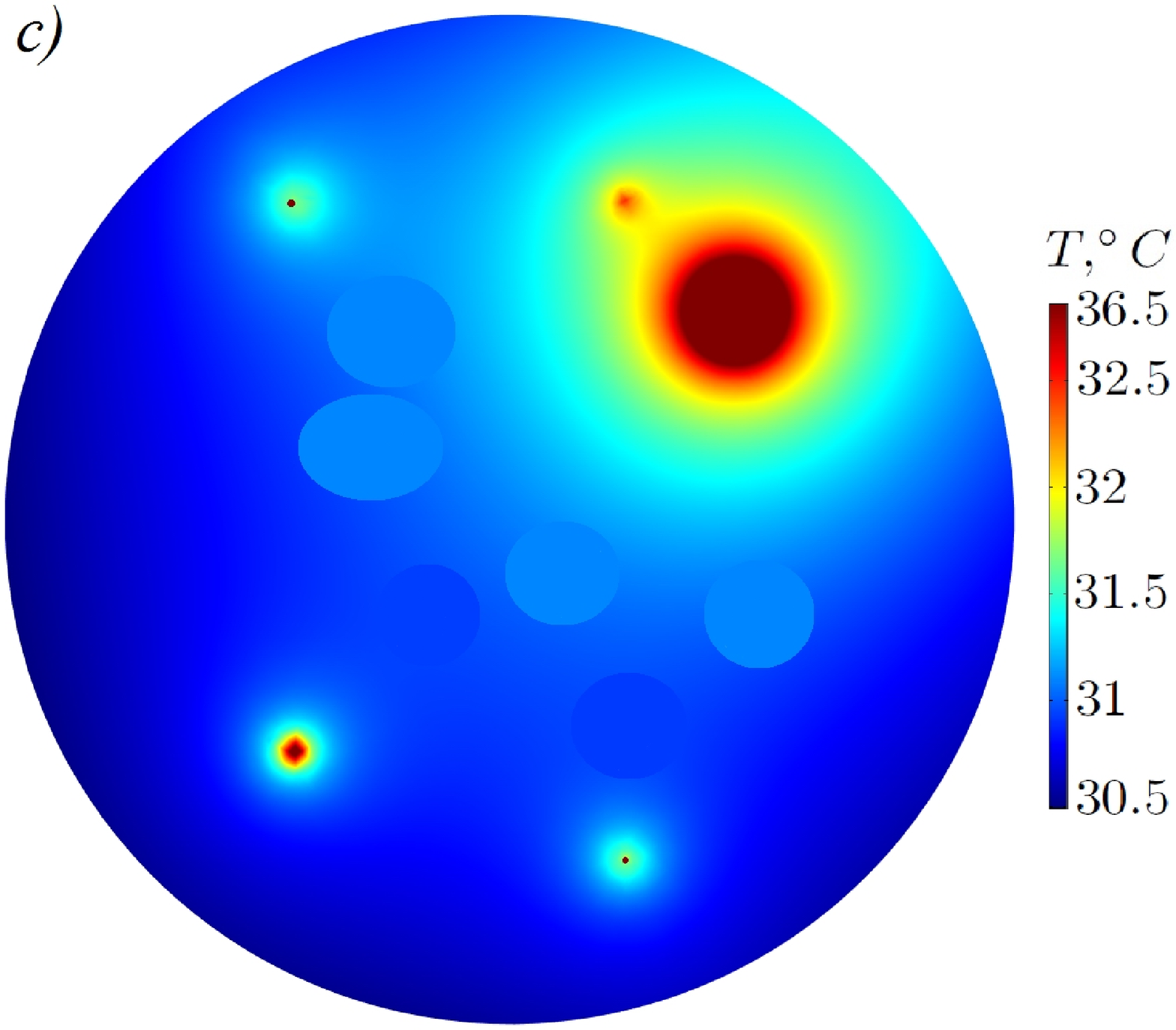}
\includegraphics[width=6cm]{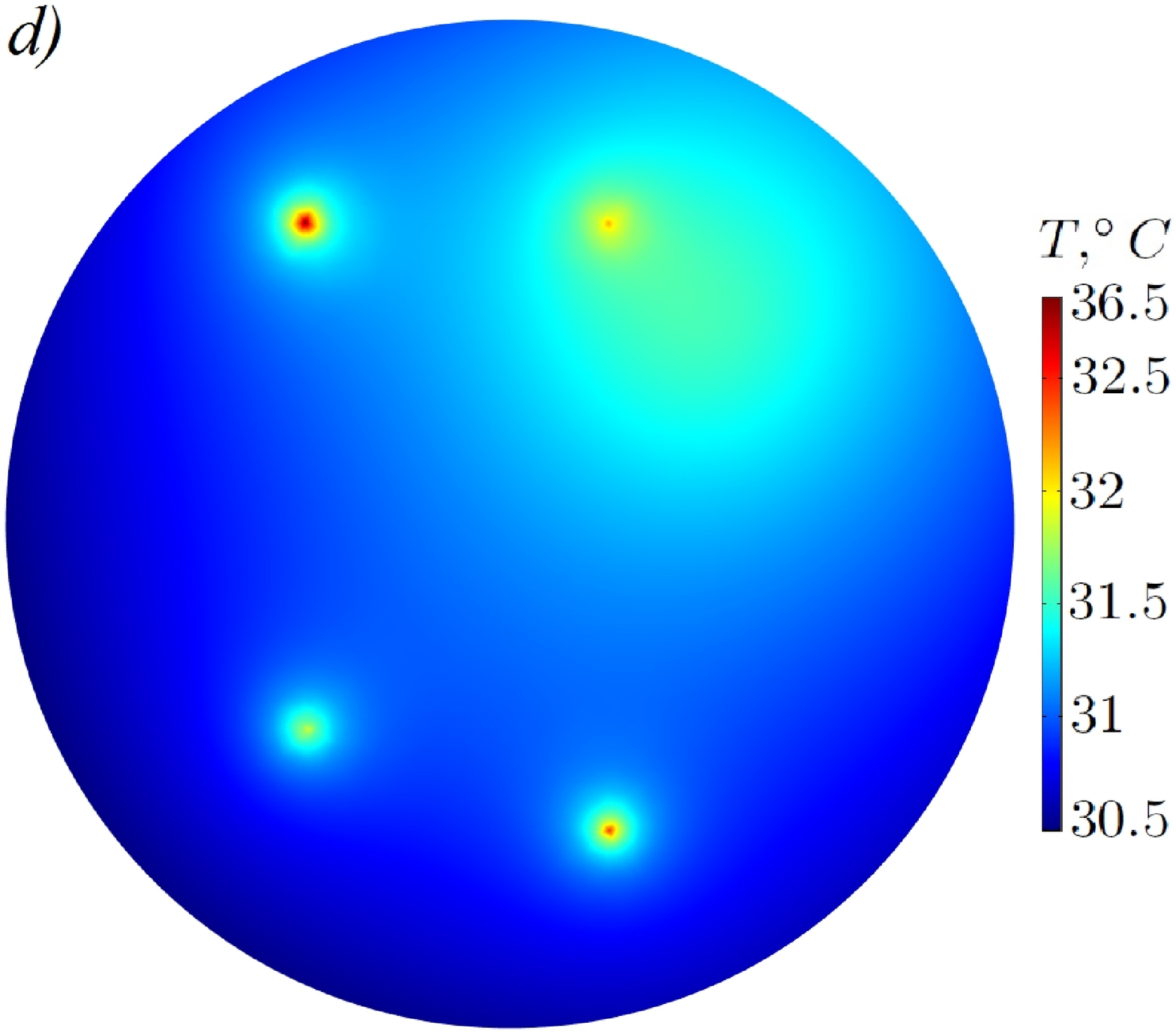}
}
\begin{center}
{\large Fig. 3. Temperature map of the model with tumor $R=0.5$\,sm, at a depth 2.5\,sm and {\it a}) $z=4$\,sm, {\it b}) $z=3$\,sm, {\it c}) $z=2.5$\,sm, {\it d}) $z=2$\,sm}
\end{center}
\label{Fig-z1234}
\end{figure}
{\large
We measure 20 patients for 25 minutes with an interval of 5 minutes.
Most medical measurements have fully confirmed the thermoregulation process.
We found a trend. The temperature starts to increase from 20th minute after the decrease.
We consider this process in our model. The amount of heat that gives blood streams to biological tissues increases from 20th minute.

We perform numerical experiments and we vary the values of physical parameters of the biocomponent in each new calculation. The results of numerical simulation are compared with the results of medical measurements (Fig. 2).
We use the patient data for the left breast cancer at the point 0 \cite{losev}. As we see, the results of the numerical experiment are consistent with the data of medical measurements.

We constructed temperature maps for the model with a tumor based on the results of numerical experiment (Fig. 3).
Strong asymmetric temperature in regions with and without a tumor is observed on the basis of this temperature map \cite{Gonzalez,novochadov}.}

\begin{figure}[!ht]
\centering{
\includegraphics[width=6cm]{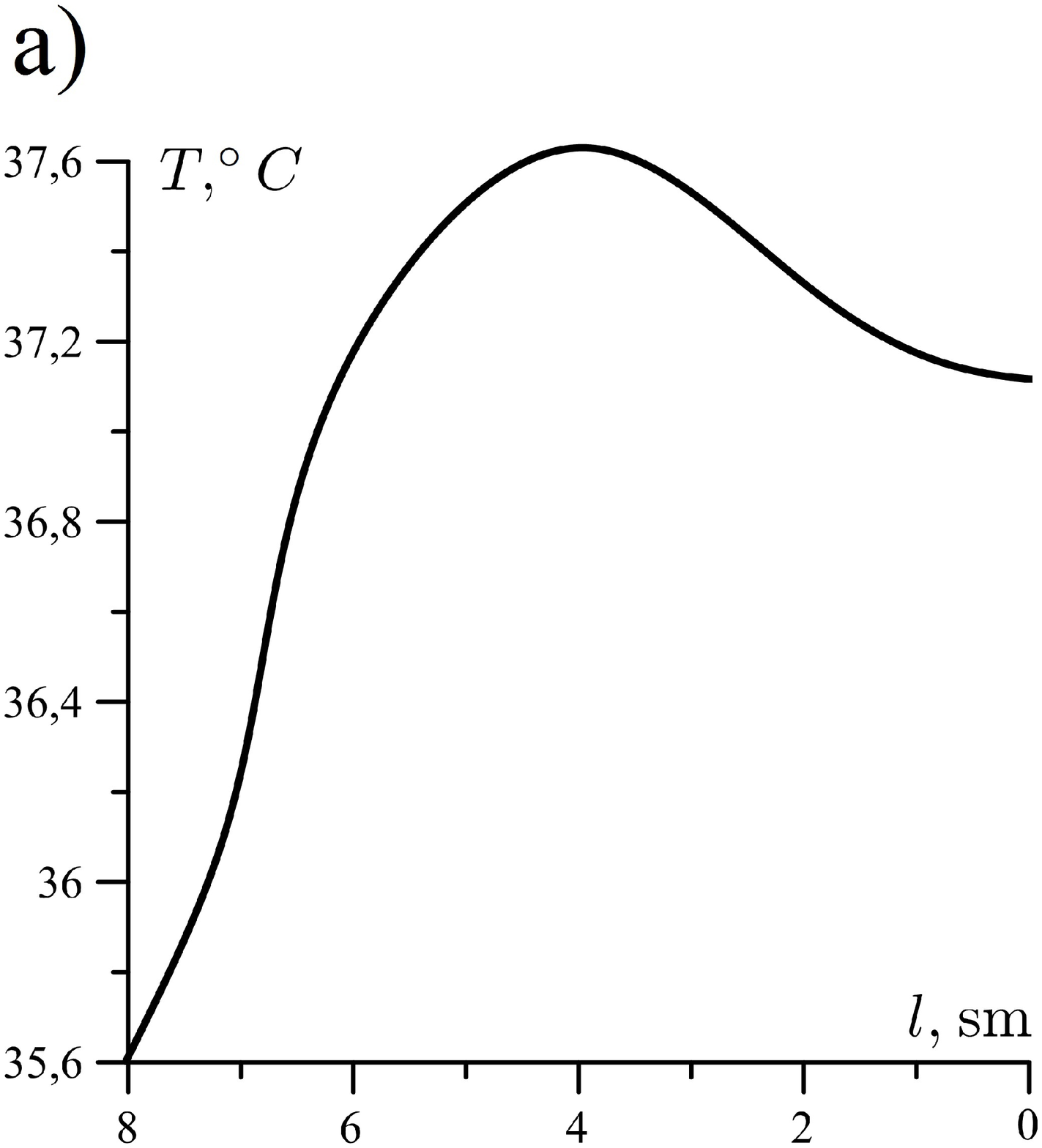}
\includegraphics[width=6cm]{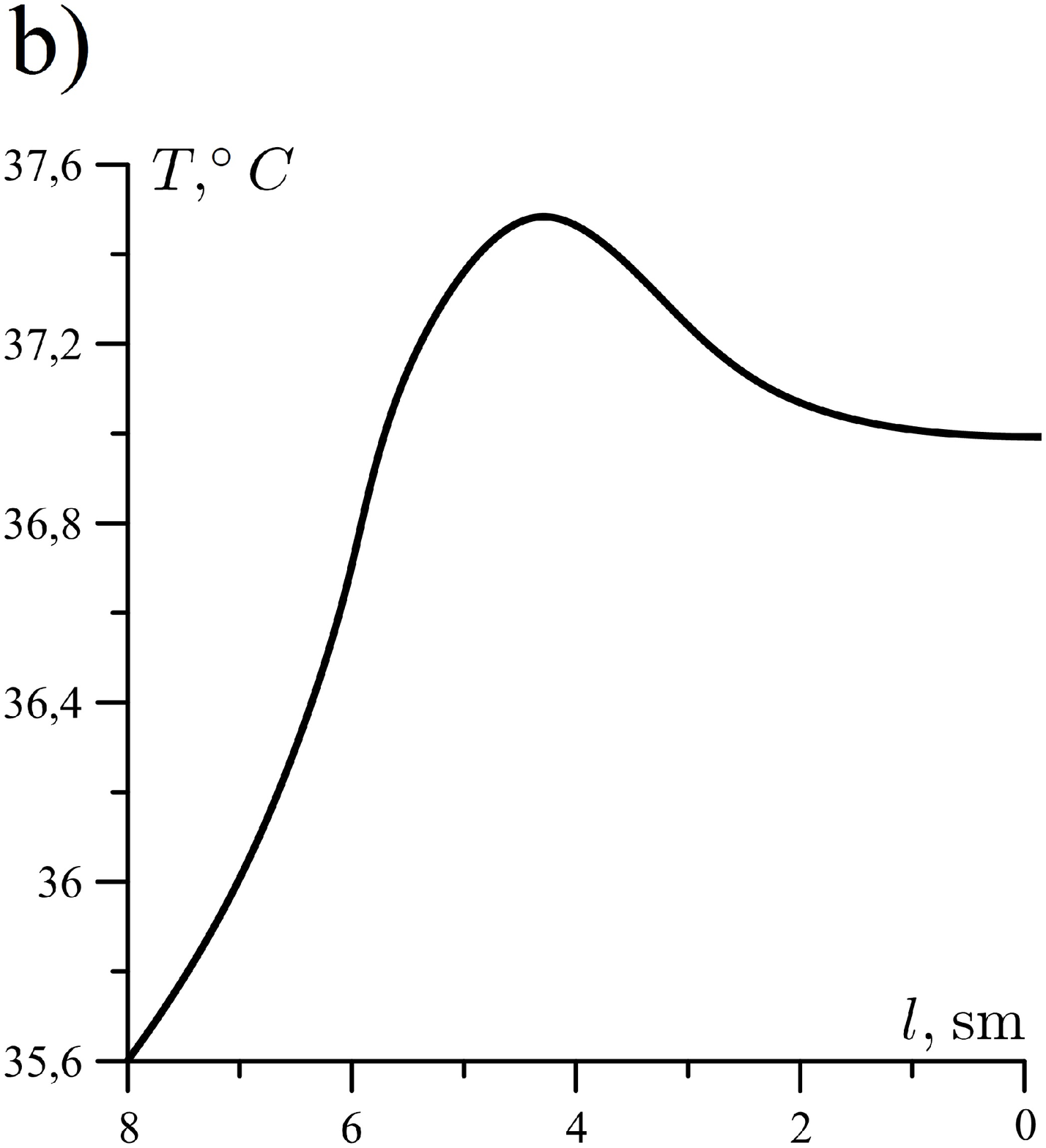}
\includegraphics[width=6cm]{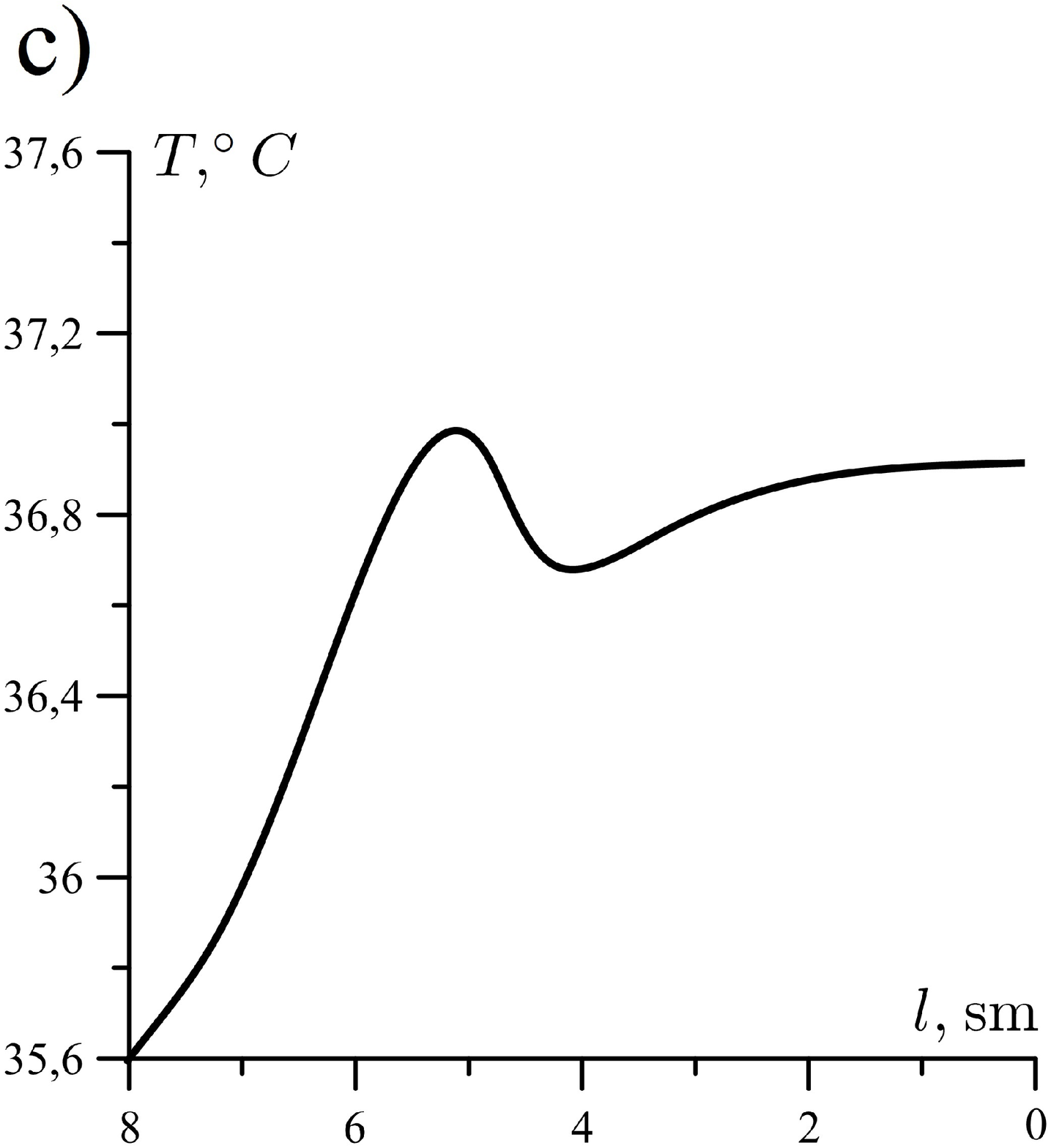}
\includegraphics[width=6cm]{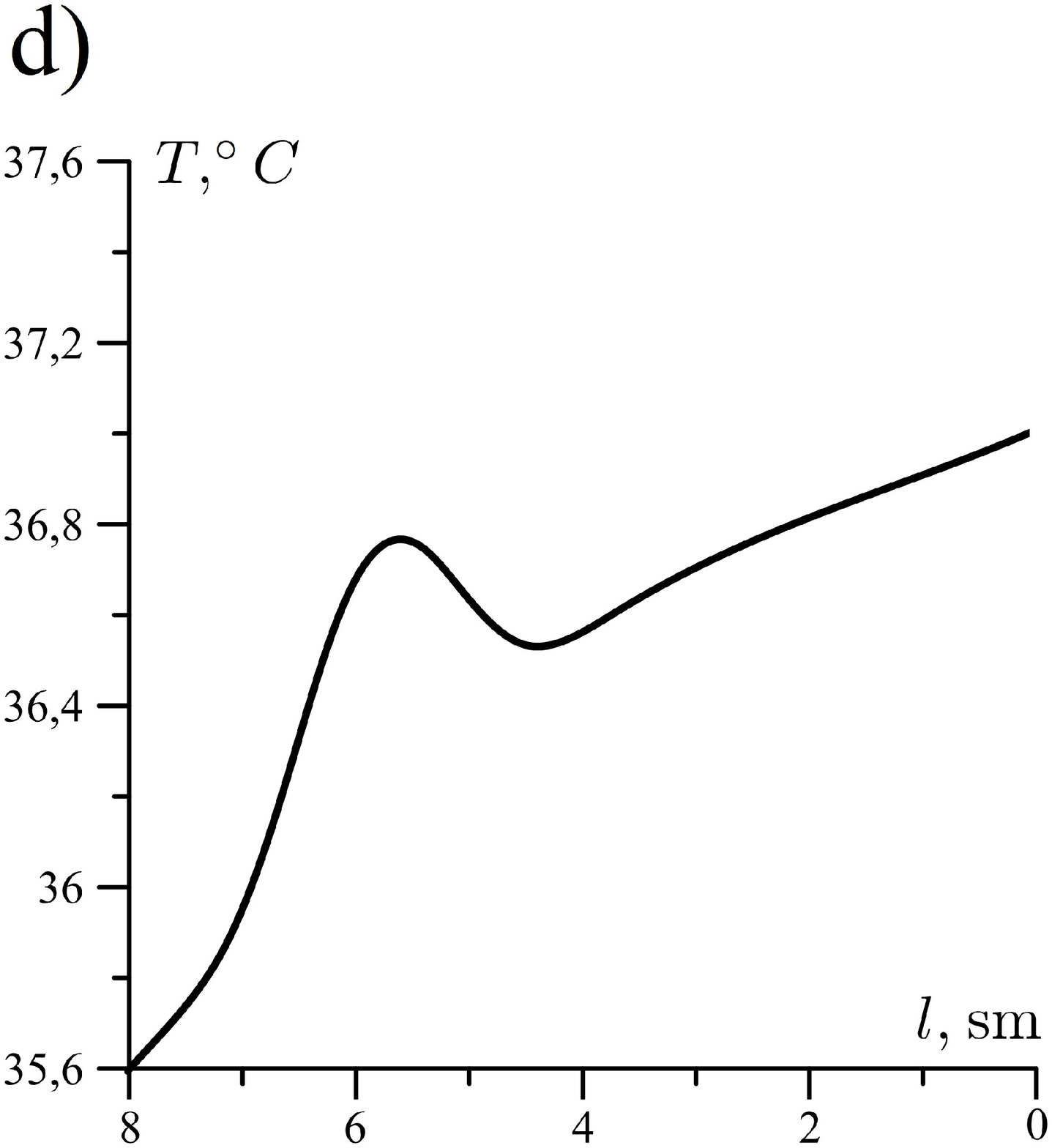}
}
\begin{center}
{\large Fig. 4. Temperature distribution along the depth for tumors of various diameters $D$, {\it a})~$D=2$\,sm, {\it b})~$D=1.5$\,sm, {\it c})~$D=1$\,sm, {\it d})~$D=0.7$\,sm}
\end{center}
\label{Fig-g1234}
\end{figure}
{\large
We carried out the computational experiments to study the dependence of temperature fields on the presence of tumor tissue in a mammary gland.
Malignant neoplasms, especially in the early stages of development, have an extremely high heat release, with respect to the remaining biological components. We examined tumors of different diameters ($D=2$\,sm, $D=1.5$\,sm, $D=1$\,sm, $D=0.7$\,sm).

We obtained the following results (Fig. 4).
The radius of a tumor affects the temperature background inside the volume of mammary gland.
The average temperature is higher, the larger the radius of the tumor. This is the main problem of diagnosis: detection of cancer at an early stage ($ R <0.5$). A tumor of this size is difficult to detect with modern diagnostic methods.}

\section{Conclusion}

{\large We developed the models for carrying out computational experiments. Our models take into account the complex structure of biotissue.

We conducted a series of computational experiments on modeling the electric field and thermodynamic temperature in the tissues of a breast. Numerical modeling of deep temperature gives an agreement with the data of medical measurements, which takes into account the multicomponent and inhomogenety of physical parameters.

Research in this area will greatly help to study and describe the physical processes in living organisms and the influence of physical processes on medical measurements.
In our case, computer simulation allows us to estimate the error of the RTM method.}

\hfill \break
\noindent{\bf Acknowledgments.}
A.V.~Khoperskov is thankful to the Ministry of Education and Science of the Russian Federation (project No.~2.852.2017/4.6).
 M.V.~Polya\-kov and T.V.~Zamechnic thanks the RFBR grant and Volgograd Region Administration (No.~15-47-02642).

\end{document}